\documentclass[a4paper,twocolumn,english,8pt,amssymb,nofootinbib,superscriptaddress]{revtex4}
\usepackage{lmodern}
\usepackage{lmodern}

\usepackage[T1]{fontenc}
\usepackage[latin9]{inputenc}
\usepackage{babel}
\usepackage{mathrsfs}
\usepackage{amsmath}
\usepackage{amssymb}
\usepackage{esint}
\usepackage[unicode=true,pdfusetitle,
bookmarks=true,bookmarksnumbered=false,bookmarksopen=false,
breaklinks=false,pdfborder={0 0 1},backref=false,colorlinks=false]
{hyperref}

\makeatletter

\@ifundefined{textcolor}{}
{%
	\definecolor{BLACK}{gray}{0}
	\definecolor{WHITE}{gray}{1}
	\definecolor{RED}{rgb}{1,0,0}
	\definecolor{GREEN}{rgb}{0,1,0}
	\definecolor{BLUE}{rgb}{0,0,1}
	\definecolor{CYAN}{cmyk}{1,0,0,0}
	\definecolor{MAGENTA}{cmyk}{0,1,0,0}
	\definecolor{YELLOW}{cmyk}{0,0,1,0}
}

\usepackage{babel}
\usepackage{babel}
\usepackage{babel}
\usepackage{babel}
\usepackage{babel}
\usepackage{babel}
\usepackage{babel}
\usepackage{babel}
\usepackage{babel}
\usepackage{babel}
\usepackage{babel}
\usepackage{babel}
\usepackage{babel}
\usepackage{babel}
\usepackage{babel}
\usepackage{babel}
\usepackage{babel}
\usepackage{graphicx}
\def\b{\begin{equation}}
\def\e{\end{equation}}

\@ifundefined{textcolor}{}{%
	\definecolor{BLACK}{gray}{0}
	\definecolor{WHITE}{gray}{1}
	\definecolor{RED}{rgb}{1,0,0}
	\definecolor{GREEN}{rgb}{0,1,0}
	\definecolor{BLUE}{rgb}{0,0,1}
	\definecolor{CYAN}{cmyk}{1,0,0,0}
	\definecolor{MAGENTA}{cmyk}{0,1,0,0}
	\definecolor{YELLOW}{cmyk}{0,0,1,0}
}

\usepackage{latexsym}\usepackage{bm}

\makeatother

\begin{document}
	\title{Light rings around five dimensional stationary black holes and naked singularities}
	
	\author{Aydin Tavlayan}
	
	\email{aydint@metu.edu.tr}
	
	\selectlanguage{english}%
	
	\affiliation{Department of Physics,\\
		Middle East Technical University, 06800 Ankara, Turkey}
	\author{Bayram Tekin}
	
	\email{btekin@metu.edu.tr}
	\affiliation{Department of Physics,\\
		Middle East Technical University, 06800 Ankara, Turkey}

	\selectlanguage{english}%
\begin{abstract}
The existence of light rings in a spacetime is closely related to the existence of black hole horizons and observables such as the ringdown and the shadow. Black holes, compared to nonvacuum ultracompact objects, have rather unique environments. To this aim, recently [P.V.P. Cunha and C.A.R. Herdeiro, Phys. Rev. Lett. \textbf{124}, 181101 (2020)] topological arguments, independent of the underlying gravity theory, were developed to prove the existence of unstable light rings outside the Killing horizon of {\it four} dimensional asymptotically flat, stationary, axisymmetric, nonextremal black holes. Here we extend these arguments to five-dimensional stationary black holes. Generically in five dimensions, there are two possible conserved angular momenta, hence the four-dimensional discussion does not extend verbatim to five dimensions; nevertheless, we prove that there is a light ring for each rotation sense for a stationary black hole. We give the static and the Myers-Perry rotating black holes as examples. We also show that when the horizon of the black hole disappears and the singularity becomes naked, only one of the light rings survives; a similar phenomenon also occurs in four dimensions which might allow testing the cosmic censorship hypothesis.
\end{abstract}
\maketitle

\section{Introduction}

Once considered as highly exotic objects that may not even exist, black holes have entered into the realm of direct observations in various ways: as sources of gravitational waves produced by their merger with each other \cite{merger,intermediate} (or with other compact objects such as neutron stars \cite{neutron}); or via the image of their environment \cite{EHT}. In both types of these observations, it is clear that the observables associated with the black holes are quite subtle: for the merger of binary black holes, the gravitational wave that hits the detector has a specific amplitude and frequency {\it variation}  over the time of observation which match the combined analytical and numerical 
predictions of general relativity (in the inspiral, merger and ringdown phases of the event). These allow one to determine the properties of the individual black holes (such as their masses) that take part in the merger; and the luminosity distance at which the event took place.  For the image of the supermassive black holes taken by the Event Horizon Telescope, one relies on the light rings, which are special bound null unstable geodesics around the black hole: basically a photon can orbit around a black hole at a constant radius along the equatorial plane like a planet rotating around a central body.  But there could in principle be ultracompact objects without horizons, made of some form of matter, that can mimic black holes. The question is  to understand the differences  in the environment of the  horizonless ultracompact objects  and black holes. 

A black hole in a vacuum, without a nontrivial environment, is hard to detect: to be clear, it certainly has a very unique gravitational skeleton ({\it i.e.} all its multipole moments are related to each other as it has no additional hair beside its mass, angular momentum, and electric charge) which can be compared to a neutron star with all different multipole moments. But, it is clear that the observables we can practically measure about a black hole are not the gravitational multipole moments. Hence one must resort to the environment of a black hole which seems to have rather unique properties (just like the mentioned unique properties of the vacuum black hole itself), such as the light rings; and these are related to the observables such as the ringdown and the shadow of the black hole. From this vantage point, the result of \cite{Cunha} becomes quite remarkable: under certain assumptions of symmetry and regularity, the existence of an unstable light ring (for each rotation sense) is related to the existence of null Killing horizons. On the other hand, ultracompact objects without horizons have a different environment as far as the light ring structure is concerned \cite{Cunha2}. Hence one can detect a black hole, by observing its environment, and in particular its light ring structure and the substructure. For generic photon orbits (not just the light rings) around the Kerr black hole \cite{Kerr}, see \cite{Teo} and \cite{Aydin} for the discussions and the references therein.

In the current work, we extend the topological arguments of \cite{Cunha} (which essentially boils down to defining a vector field that vanishes only at the location of the light ring) to generic five-dimensional stationary black holes. To understand the light ring structure around these black holes, as it will be clear, the effective potential for the light has to be chosen carefully since there are two conserved angular momenta in five dimensions, and the impact parameters of light enter into the effective potential functions. To properly set the stage, in Sec. II, we start with the static five-dimensional spacetime and give the Schwarzschild-Tangherlini black hole as an example. In Sec. III, we introduce the topological techniques and apply them to the static black hole. In Sec. IV, a generic stationary black hole is studied: first we discuss the degenerate angular momenta case and then move on to the distinct angular momenta case. As an example, we also study the Myers-Perry rotating solution. Section V is devoted to a brief study of the case when the event horizon disappears and one is left with a rotating, massive naked singularity. We show that in this case, in four and five dimensions, there is a single light ring.

\section{Five dimensional Static Spacetimes}

In the coordinates $(t,r,\theta,\phi_1,\phi_2)$ \cite{MP}, with the ranges,
\begin{equation}
\begin{aligned}
&t \in (-\infty, \infty), \,\,\,\,\,
r \in[r_{H}, \infty), \\
&\theta \in\left[0, \frac{\pi}{2}\right], \,\,\,\,\,
\varphi_{1} \in[0,2 \pi], \,\,\,\,\,
\varphi_{2} \in[0,2 \pi],
\end{aligned}
\end{equation}
the metric of the $4+1$ dimensional static black hole spacetime can be written as 
\begin{eqnarray}
ds^2&=&\xi^2dt^2+g_{rr}\left(r,\theta\right)dr^2+g_{\theta\theta}\left(r,\theta\right)d\theta^2\nonumber\\
&&+\eta_{1}^2d\phi_1^2+\eta_{2}^2d\phi_2^2,
\end{eqnarray}
where the timelike Killing vector field reads as $\xi := \frac{\partial}{\partial t}$ with the norm $\xi^2 = g_{tt}\left(r,\theta\right)$, and the rotation Killing vectors read as 
$\eta_{1} := \frac{\partial}{\partial \phi_1}$ and $\eta_{2} := \frac{\partial}{\partial \phi_2}$ with the obvious norms $\eta_{1}^2 =g_{\phi_1\phi_1}\left(r,\theta\right)$ etc. So $(t,\phi_1,\phi_2)$ are Killing coordinates, and $(r,\theta)$ are essential coordinates. We assume that there is a Killing horizon $r_H>0$ at which $\xi^2(r_{H})=0$ and for all $r> r_H$, we have $\xi^2(r) <0$. We also assume asymptotic flatness and causality $\xi^2(r \rightarrow\infty) \rightarrow -1$ and  $\eta_{1,2}$ are spacelike. 

In this background, we are interested in the existence of bound null geodesics, and in particular light rings. One way to do is to study the null Hamiltonian condition for a photon, which states that
\begin{equation}
H=\frac{1}{2}g^{\mu\nu}p_{\mu}p_{\nu}=0,
\end{equation}
where $p^{\mu}= \frac{dx^\mu}{d \lambda}$ represents the momentum of the photon where $\lambda$ is an affine parameter. The Killing symmetries, dictate the following conserved quantities 
\begin{eqnarray}
&&p_t= \langle \xi,p\rangle=:-E, \nonumber\\
&&p_{\phi_1}=\langle \eta_{1},p\rangle=:\Phi_1, \nonumber\\
&&p_{\phi_2}=\langle \eta_{2},p\rangle=:\Phi_2,
\end{eqnarray}
where $E$, $\Phi_1$, and $\Phi_2$ represent the energy and angular momenta of the photon at spatial infinity, respectively.  

We can split the Hamiltonian into a kinetic and a potential part as
\begin{eqnarray}
&&K=g^{rr}p_r^2+g^{\theta\theta}p_{\theta}^2, \nonumber\\
&&V=g^{tt}E^2+g^{\phi_1\phi_1}\Phi_1^2+g^{\phi_2\phi_2}\Phi_2^2.
\end{eqnarray}
If we restrict to the light ring, we have 
\begin{equation}
p_r=p_{\theta}=\dot{p}_{\mu}=0, \label{null momenta}
\end{equation}
which implies that $K=0$ and therefore $V=0$. This is the {\it first} light ring condition. The second light ring condition follows from the Hamilton's equations which state that
\begin{eqnarray}
\dot{p}_{\mu}&=&-\partial_{\mu}\left(\frac{1}{2}g^{\alpha\beta}p_{\alpha}p_{\beta}\right)\nonumber\\
&=&-\frac{1}{2}\left(\partial_{\mu}g^{rr}p_r^2+\partial_{\mu}g^{\theta\theta}p_{\theta}^2+\partial_{\mu}V\right). \label{null momenta2}
\end{eqnarray}
The equation (\ref{null momenta2}) together with (\ref{null momenta}) implies that $\partial_{\mu}V=0$ which is the {\it second} light ring condition. 

The crux of the above argument is this: one can study the light rings by just looking at the 
potential term only. But the drawback this potential is that it directly depends on the parameters of the photon, $E$, $\Phi_1$, and $\Phi_2$. We would like to separate the properties of the photon from the properties of the background spacetime. In order to get rid of this dependence, it is useful to write the potential in the following form: first let us define
\begin{equation}
D:=-\xi^2 \eta_1^2\eta_2^2
\end{equation}
and multiply it with the potential
\begin{eqnarray}
-V\times D =\eta_1^2\eta_2^2E^2+\xi^2 \eta_2^2\Phi_1^2 +\xi^2 \eta_1^2\Phi_2^2. \label{pot}
\end{eqnarray}
At this stage, to better understand the problem, let us take the angular momenta to be equal (a simplification which we shall remove in the spinning black hole case). Then $\Phi_1=\Phi_2 :=\Phi$. Furthermore, due to the equivalence principle, there is no gravitational rainbow (photons with different energies can circle the same geodesic), therefore we can eliminate the energy of the photon by using the inverse impact parameter as usual\footnote{Here, we assume that the photon is sufficiently energetic so that its wavelength is small compared to the variations in the gravitational field, hence the ray optics approximation works.}
\begin{equation}
\sigma:=\frac{E}{\Phi},
\end{equation}
with the help of which, the effective potential factors as
\begin{equation}
V=\frac{\Phi^2}{\xi^2}\left(\sigma-\sigma_-\right)\left(\sigma-\sigma_+\right)
\end{equation}
where the effective potential functions are independent of the properties of the light and are determined by the geometry alone:
\begin{equation}
\sigma_{\pm}=\pm \sqrt{-\xi^2\left(\frac{1}{\eta_1^2}+\frac{1}{\eta_2^2}\right)}.
\end{equation}
Now, the first light ring condition ($ V=0$) states that $\sigma$ is either equal to $\sigma_-$ or $\sigma_+$, and this only determines the impact parameter in terms of the geometry. But, the second light ring condition, that is the flatness of the potential in {\it all} directions $\partial_{\mu}V=0$, carries a great deal more information which we shall explore now.

In order to understand the 'gradient flows' associated with the second light ring condition, it is best to define a two dimensional vector field as  \cite{Cunha2}
\begin{equation}
v_r :=\frac{\partial_r\sigma_{\pm}}{\sqrt{g_{rr}}}, \hspace{0.5 cm} v_{\theta}:=\frac{\partial_{\theta}\sigma_{\pm}}{\sqrt{g_{\theta\theta}}},
\end{equation}
which was normalized as above to yield $ \partial^\mu\sigma_{\pm} \partial_\mu \sigma_{\pm}= v_r^2+ v_\theta^2 =:v^2$ Hence the second light ring condition dictates that $\vec{v}=0$. Furthermore, defining the angle $\Omega$ as $v_r= v \cos\Omega$ and $v_\theta= v\sin\Omega$, one can see that the integral $\oint_C d \Omega$ over a closed curve $C$ in the $(r,\theta)$ space should yield $2 \pi w$ with $w$ being the winding number taking values in integers. Hence as shown in \cite{Cunha}, $w$ is a well-defined topological number that one can assign to light rings:
\begin{equation}
w = \frac{1}{2 \pi} \oint_C d \Omega(r,\theta), \label{winding}
\end{equation}
where $C$ can be deformed to any other contour as long as a light ring is not crossed. The sign of $w$ is just a convention: for a counterclockwise $C$, negative winding corresponds to a standard light ring.

At this stage, let us consider the simplest case: the five dimensional Schwarzschild-Tangherlini black hole with the metric functions given as
\begin{eqnarray}
&&g_{t t}=-\left(1-\frac{\mu}{r^{2}}\right),\,\,\,\, g_{r r}=\left(1-\frac{\mu}{r^{2}}\right)^{-1}\nonumber\\
&&g_{\theta \theta}=r^{2},\,\,\,\,  g_{\phi_{1} \phi_{1}}=r^{2} \sin ^{2} \theta,\nonumber\\
&&g_{\phi_{2} \phi_{2}}=r^{2} \cos ^{2}, \theta\nonumber\\
&&D=\left(r^{2}-\mu\right)r^{2} \sin ^{2} \theta \cos ^{2} \theta,
\end{eqnarray}
where $\mu$ is related to the mass of the black hole  via
\begin{equation}
\mu=\frac{8 G M}{3 \pi}.
\end{equation}
Hence, the effective potential functions become
\begin{eqnarray}
&&\sigma_{\pm}=\frac{\pm 1}{r^{2} \sin \theta \cos \theta} \sqrt{r^{2}-\mu}.
\end{eqnarray}
The vector field components can be calculated as 
\begin{equation}
v_r=\pm \frac{1}{r^4 \sin \theta \cos \theta}\left(2\mu-r^2\right),
\label{vr}
\end{equation}
and
\begin{equation}
v_{\theta}=\mp \frac{4 \sqrt{r^2-\mu}}{r^3}\left(\frac{\cos 2\theta}{\sin^2 2\theta}\right).
\label{vtheta}
\end{equation}
As stated above, the light ring corresponds to the particular point in the vector field which has the property
\begin{equation}
v_r^2+v_{\theta}^2=0 \rightarrow v=0.
\end{equation}
Therefore, we can conclude that the light ring is located at $r=\sqrt{2\mu}$ and $\theta=\frac{\pi}{4}$. In other words, we have a standard light ring outside the horizon for a five dimensional static black hole. Now, we can confirm this by using the topological charge and the winding number concepts. 

\section{Contour Analysis and the Winding Number for the static black hole in 5 dimensions}
\begin{figure}
	\centering
	\includegraphics[width=1.0\linewidth]{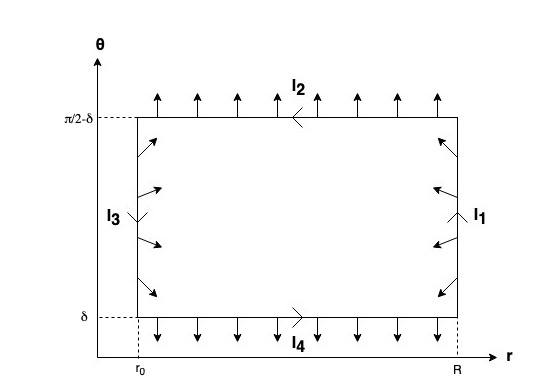}
	\caption{The representation of the results found in the contour analysis. The vector field, $\vec{v}$, obtained by using the effective potential of the positive rotation sense can be seen along the contour. The full negative winding is apparent.}
	\label{fig:i1}
\end{figure}

Let us assume, we have a contour outside the horizon which can be described as in Fig. (\ref{fig:i1}) with the following line segments

\vskip 0.5 cm

$I_1$: $r=R$,\hskip 0.3 cm $\delta\le \theta \le \frac{\pi}{2}-\delta$,
\vskip 0.5 cm
$I_2$: $\theta=\frac{\pi}{2}-\delta$, \hskip 0.3 cm$r_0\le r \le R$,
\vskip 0.5 cm
$I_3$: $r=r_0$, \hskip 0.3 cm$\delta\le \theta \le \frac{\pi}{2}-\delta$,
\vskip 0.5 cm
$I_4$: $\theta=\delta$, \hskip 0.3 cm$r_0\le r \le R$.		
\vskip 0.5 cm
Now, we would like to investigate how the vector field $\vec{v}$ changes along this contour. 
But we must cover all the exterior region to the black hole which means at the end, our contour should be extended to all the $(r,\theta)$ plane outside the horizon. Hence we must take the limits $\delta \rightarrow 0$, $r_0 \rightarrow r_H$ and $R\rightarrow \infty$. The order of the limits is important \cite{Cunha}. Let us study each line segment separately.

\subsubsection{Line segment $I_4$}

From (\ref{vr}) and (\ref{vtheta}), one observes that as $\delta \rightarrow 0$ and therefore $\theta \rightarrow 0$ along the line segment $I_4$: the components of the vector field become
\begin{equation}
v_r \propto \pm \frac{1}{\sin \theta}
\end{equation}
and 
\begin{equation}
v_{\theta} \propto \mp \frac{1}{\sin^2 2\theta}.
\end{equation}
Hence, $v_{\theta}$ is the dominant component of the vector field and we have
\begin{equation}
\Omega=\left.\arcsin \left(\frac{v_{\theta}}{v}\right)\right|_{0} \rightarrow \mp \pi / 2  \text { for } \theta \rightarrow 0
\end{equation}
along $I_4$.

\subsubsection{Line segment $I_2$}
Similarly, as $\delta \rightarrow 0$ and therefore $\theta \rightarrow \frac{\pi}{2}$ along the line segment $I_2$
\begin{equation}
v_r \propto \pm \frac{1}{\cos \theta}
\end{equation}
and 
\begin{equation}
v_{\theta} \propto \mp \frac{1}{\sin^2 2\theta}.
\end{equation}
Hence, $v_{\theta}$ is  the dominant component of the vector field and 
\begin{equation}
\Omega=\left.\arcsin \left(\frac{v_{\theta}}{v}\right)\right|_{\frac{\pi}{2}} \rightarrow \pm \pi / 2  \text { for } \theta \rightarrow \frac{\pi}{2}
\end{equation}
along $I_2$.

\subsubsection{Line segment $I_3$}

\noindent The event horizon for the five dimensional static black hole in the given coordinates is located at $r_H=\sqrt{\mu}$. While approaching the horizon, the $r$ component of the vector field  (\ref{vr}) does not change sign along  $I_3$. The term in the parenthesis is always positive while approaching the event horizon and $r^4 \sin \theta \cos \theta = \frac{r^4}{2} \sin 2\theta$ is always positive for $\theta \in \left[0,\frac{\pi}{2}\right]$. Therefore, for
\begin{equation}
\sigma=\sigma_+,  \hspace{0.5 cm}  v_r \rightarrow +
\end{equation}
and for
\begin{equation}
\sigma=\sigma_-,  \hspace{0.5 cm}  v_r \rightarrow -.
\end{equation}
Yet, the $\theta$ component of the vector field changes (\ref{vtheta}) sign along  $I_3$, because it has $\cos 2\theta$, which is negative for $\theta \in \left[\frac{\pi}{4},\frac{\pi}{2} \right]$ and is positive for $\theta \in \left[0,\frac{\pi}{4}\right]$. This constitutes the half of the winding as can be seen in (\ref{fig:i1}).

\subsubsection{Line segment $I_1$}

The same argument with the line segment $I_3$ can be used here. The $v_r$ does not change sign along $I_1$ and points opposite to $I_3$. The $v_\theta$ component of the vector field changes sign because for a counterclockwise rotation, it starts at a negative direction and ends in the positive direction for the positive rotation sense and this constitutes the other half of the winding (\ref{fig:i1}).

To summarize, in order to evaluate (\ref{winding}), we decomposed the contour into four lines, and investigated each lines separately. In other words, we wrote the winding number as (where the proper limits are to be understood)
\begin{equation}
\omega:=\omega_{I_1}+\omega_{I_2}+\omega_{I_3}+\omega_{I_4},
\end{equation}
where
\begin{eqnarray}
\omega_{I_i}&:=\frac{1}{2\pi}&\int_{I_i} d \Omega(r,\theta), \,\,\, i \in (1,2,3,4).\label{windingC}
\end{eqnarray}
We showed that there is no contribution to the winding number along the lines $I_{2}$ and $I_{4}$, in other words, $\omega_{I_2}=0$ and $\omega_{I_4}=0$. We also showed that there are two negative half windings along the lines $I_{1}$ and $I_{3}$, and hence $\omega_{I_1}=-\frac{1}{2}$ and $\omega_{I_3}=-\frac{1}{2}$. In conclusion, we obtained $\omega=-1$, which implies that we had a standard light ring inside the contour.

\begin{figure}
	\centering
	\includegraphics[width=0.8\linewidth]{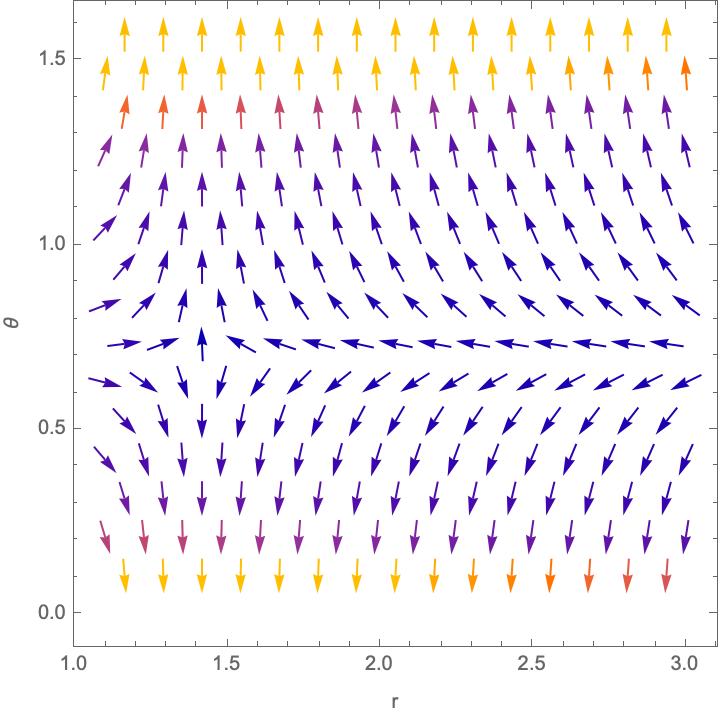}
	\caption{$\vec{v}= \left(\frac{1}{r^4 \sin \theta \cos \theta}\left(2-r^2\right),-\frac{4 \sqrt{r^2-1}}{r^3}\left(\frac{\cos 2\theta}{\sin^2 2\theta}\right)\right)$ obtained by using the effective potential function associated with the positive rotation sense is plotted in neighborhood of the standard light ring ($r= \sqrt{2}$, $\theta= \pi/4$) for the static spacetime with two equal angular momenta.}
	\label{fig:staticcase}
\end{figure}

\begin{figure}
	\centering
	\includegraphics[width=0.8\linewidth]{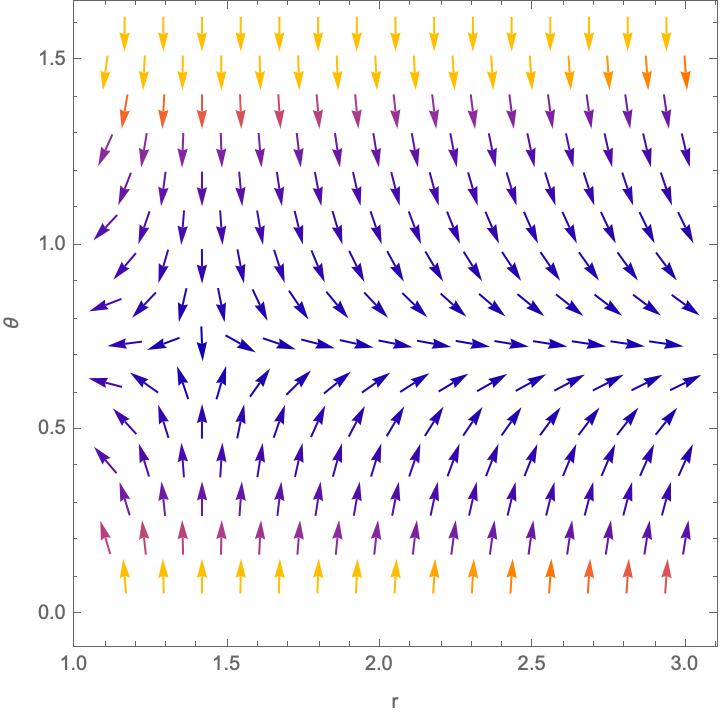}
	\caption{$\vec{v}= \left(\frac{-1}{r^4 \sin \theta \cos \theta}\left(2-r^2\right),\frac{4 \sqrt{r^2-1}}{r^3}\left(\frac{\cos 2\theta}{\sin^2 2\theta}\right)\right)$ obtained by using the effective potential function associated with the negative rotation sense is plotted in neighborhood of the standard light ring ($r= \sqrt{2}$, $\theta= \pi/4$) for the static spacetime with two equal angular momenta.}
	\label{fig:staticcaseneg}
\end{figure}

The general behavior of the vector field around the light ring is plotted in Fig.
(\ref{fig:staticcase}) and Fig. (\ref{fig:staticcaseneg}).


\section{Spinning Black Holes in 5 dimensions}

Let us now consider three Killing vectors to be nonorthogonal to each other, such that in the
 coordinates $(t,r,\theta,\phi_1,\phi_2)$ the metric of the five dimensional stationary black hole reads as 
\begin{eqnarray}
ds^2&=&\xi^2dt^2+g_{rr}\left(r,\theta\right)dr^2+g_{\theta\theta}\left(r,\theta\right)d\theta^2\nonumber\\
&&+\eta_{1}^2d\phi_1^2+\eta_{2}^2d\phi_2^2 +2(\xi,\eta_{2})dt d\phi_{2}\nonumber\\
&&+2(\eta_{1},\eta_{2})d\phi_{1}d\phi_{2}+2(\xi,\eta_{1})dt d\phi_{1},
\end{eqnarray}
where $(\eta_{1},\eta_{2})=g_{\phi_1\phi_2}\left(r,\theta\right)$ etc.
As in the static case, the Hamiltonian can be split into the kinetic and potential parts. The kinetic part is the same as the static case,
\begin{eqnarray}
&&K=g^{rr}p_r^2+g^{\theta\theta}p_{\theta}^2, 
\end{eqnarray}
but the potential now has cross terms and reads as 
\begin{eqnarray}
V&=&g^{tt}E^2+g^{\phi_1\phi_1}\Phi_1^2+g^{\phi_2\phi_2}\Phi_2^2\nonumber\\
&&-2g^{t\phi_1} E \Phi_1-2g^{t\phi_2} E \Phi_2 + 2g^{\phi_1\phi_2} \Phi_1 \Phi_2,
\end{eqnarray}
which can be recast as
\begin{eqnarray}
V&=&-\frac{1}{D}\left[E^2\left(\eta_{1}^2 \eta_{2}^2-(\eta_{1},\eta_{2})^2\right)\right.\nonumber\\
&&\left.+\Phi_1^2\left(\xi^2 \eta_{2}^2-(\xi,\eta_{2})^2\right)\right.\nonumber\\
&&\left.+\Phi_2^2\left(\xi^2 \eta_{1}^2-(\xi,\eta_{1})^2\right)\right.\nonumber\\
&&\left.+2\Phi_1\Phi_2\left((\xi,\eta_{1}) (\xi,\eta_{2})-\xi^2 (\eta_{1},\eta_{2})\right)\right.\nonumber\\
&&\left.+2E\Phi_1\left((\xi,\eta_{1}) \eta_{2}^2 -(\xi,\eta_{2})(\eta_{1},\eta_{2})\right)\right.\nonumber\\
&&\left.+2E\Phi_2\left((\xi,\eta_{2}) \eta_{1}^2-(\xi,\eta_{1}) (\eta_{1},\eta_{2})\right)\right],
\label{pot3}
\end{eqnarray}
where
\begin{eqnarray}
D&:=&\eta_{1}^2 (\xi,\eta_{2})^2 + \eta_{2}^2 (\xi,\eta_{1})^2 + \xi^2 (\eta_{1},\eta_{2})^2 \nonumber\\
&&- \xi^2 \eta_{1}^2 \eta_{2}^2 - 2 (\xi,\eta_{1}) (\xi,\eta_{2}) (\eta_{1},\eta_{2}).
\end{eqnarray}

Once again, due the equivalence principle, energy of the light will not play a role by itself, and we can pull it out as a factor in the effective potential.  At this stage the discussion bifurcates: if the angular momenta of the light are equal to each other, the problem is easier to handle, if not, one needs to work a little harder. Let us study these two cases separately.

\subsection{Case I: Equal angular momenta }

Let  $\Phi_1=\Phi_2=\Phi$ and define, as in the static case, the inverse impact parameter 
\begin{equation}
\sigma := \frac{E}{\Phi}
\end{equation}
then (\ref{pot3}) reduces to
\begin{eqnarray}
&&V=-\frac{\Phi^2}{D}\left[\sigma^2\left(g_{\phi_1 \phi_1}g_{\phi_2 \phi_2}-g_{\phi_1 \phi_2}^2\right)\right. \\
&&\left.+2\sigma\left(g_{t\phi_1}\left(g_{\phi_2 \phi_2}-g_{\phi_1 \phi_2}\right)+g_{t\phi_1}\left(g_{\phi_2 \phi_2}-g_{\phi_1 \phi_2}\right)\right)\right.\nonumber\\
&&\left.\left(g_{tt}\left(g_{\phi_1 \phi_1}+g_{\phi_2 \phi_2}-2g_{\phi_1 \phi_2}\right)-\left(g_{t \phi_1}-g_{t \phi_2}\right)^2\right)\right]\nonumber
\end{eqnarray}
where, we have introduced the metric components explicitly. The effective potential factors as
\begin{equation}
V=-\frac{\Phi^2}{D}\left(\sigma-\sigma_+\right)\left(\sigma-\sigma_-\right)
\end{equation}
with
\begin{equation}
\sigma_{\pm}:=\frac{A_\pm}{B}
\end{equation}
where
\begin{eqnarray}
A_\pm:=&&-g_{\phi_2 \phi_2}g_{t \phi_1}-g_{\phi_1 \phi_1}g_{t \phi_2}\nonumber\\
&&+g_{t \phi_1}g_{\phi_1 \phi_2}+g_{t \phi_2}g_{\phi_1 \phi_2}\nonumber\\
&&\pm \sqrt{D\left(g_{\phi_1 \phi_1}+g_{\phi_2 \phi_2}-2g_{\phi_1 \phi_2}\right)}
\end{eqnarray}
and
\begin{equation}
B:=g_{\phi_1 \phi_1}g_{\phi_2 \phi_2}-g_{\phi_1 \phi_2}^2.
\end{equation}
In order to find the light ring outside the event horizon, we are going to follow a similar approach with the static case and investigate the gradient flows along a contour and take limits to cover all the space outside the Killing horizon.
\subsubsection{Axis limit}

{\bf A: {$\theta \rightarrow 0$}}

In order to understand the behavior of the metric components while approaching the axis, it is beneficial to introduce a local coordinate \cite{Cunha}
\begin{equation}
\rho^2 :=g_{\phi_1 \phi_1}
\end{equation}
and $\rho \rightarrow 0$ as $\theta \rightarrow 0$. The other metric components can be expanded in terms of $\rho$ as $\rho \rightarrow 0$. Keeping the dominant terms, we have 
\begin{eqnarray}
&&g_{tt} \approx g_{tt}^{(0)}+O\left(\rho\right),\nonumber\\
&&g_{t \phi_1} \approx \rho^{n}+O\left(\rho^{n+1}\right),\nonumber\\
&&g_{t \phi_2} \approx g_{t \phi_2}^{(0)}+O\left(\rho\right),\nonumber\\
&&g_{\phi_1 \phi_2} \approx \rho^{m}+O\left(\rho^{m+1}\right),\nonumber\\
&&g_{\phi_2 \phi_2} \approx g_{\phi_2 \phi_2}^{(0)}+O\left(\rho\right).
\end{eqnarray}
At this point, it is important to emphasize the fact that $n\ge2$ and $m\ge2$. This can easily be shown by extending the regularity ideas developed in \cite{Cunha} and require a nonvanishing scalar curvature . [We give the details of this in the Appendix.] By using these expansions, one has
\begin{eqnarray}
D&\approx&\rho^2\left(\left(g_{t \phi_2}^{\left(0\right)}\right)^2 + g_{\phi_2 \phi_2}^{\left(0\right)} \frac{\rho^{2n}}{\rho^2} + g_{tt}^{\left(0\right)} \frac{\rho^{2m}}{\rho^2} \right. \nonumber\\
&&\left.- g_{tt}^{\left(0\right)} g_{\phi_2 \phi_2}^{\left(0\right)} - 2 g_{t \phi_{2}}^{\left(0\right)} \frac{\rho^{n+m}}{\rho^2} \right).
\end{eqnarray}
Because of the fact that $g_{\phi_1 \phi_1}$ cannot go to zero faster than $g_{t \phi_1}$ and $g_{\phi_1 \phi_2}$, one has
\begin{equation}
D \approx \rho^2 \left(\left(g_{t \phi_2}^{\left(0\right)}\right)^2- g_{tt}^{\left(0\right)} g_{\phi_2 \phi_2}^{\left(0\right)}\right),
\end{equation}
while 
\begin{eqnarray}
A\approx&&\rho^2\left(-g_{t \phi_2}^{\left(0\right)}\pm\left[\left(\left(g_{t \phi_2}^{\left(0\right)}\right)^2- g_{tt}^{\left(0\right)} g_{\phi_2 \phi_2}^{\left(0\right)}\right)\right.\right.\nonumber\\
&&\left.\left.\left(1+\frac{g_{\phi_2 \phi_2}^{\left(0\right)}}{\rho^2}\right)\right]^{\frac{1}{2}}\right),
\end{eqnarray}
and
\begin{eqnarray}
B\approx&&\rho^2 g_{\phi_2 \phi_2}^{\left(0\right)}-\rho^{2m}\approx\rho^2\left(g_{\phi_2 \phi_2}^{\left(0\right)}\right).
\end{eqnarray}
As a consequence,
\begin{eqnarray}
&&\sigma_{\pm}\approx-\frac{g_{t \phi_2}^{\left(0\right)}}{g_{\phi_2 \phi_2}^{\left(0\right)}}\nonumber\\
&&\pm\frac{\left[\left(\left(g_{t \phi_2}^{\left(0\right)}\right)^2- g_{tt}^{\left(0\right)} g_{\phi_2 \phi_2}^{\left(0\right)}\right)\left(1+\frac{g_{\phi_2 \phi_2}^{\left(0\right)}}{\rho^2}\right)\right]^{\frac{1}{2}}}{g_{\phi_2 \phi_2}^{\left(0\right)}}\nonumber\\
&&\approx-\frac{g_{t \phi_2}^{\left(0\right)}}{g_{\phi_2 \phi_2}^{\left(0\right)}}\nonumber\\
&&\pm\frac{1}{\rho}\frac{\left[\left(\left(g_{t \phi_2}^{\left(0\right)}\right)^2- g_{tt}^{\left(0\right)} g_{\phi_2 \phi_2}^{\left(0\right)}\right)\left(\rho^2+g_{\phi_2 \phi_2}^{\left(0\right)}\right)\right]^{\frac{1}{2}}}{g_{\phi_2 \phi_2}^{\left(0\right)}}\nonumber\\
&&\approx \text{constant}\pm\frac{1}{\rho}\times\text{constant}.
\end{eqnarray}
The forms of $\sigma_{\pm}$ imply that $v_r\propto\frac{1}{\rho}$ and $v_{\theta}\propto\frac{1}{\rho^2}$. Therefore, $v_{\theta}$  is dominant as $\rho\rightarrow 0$. As a conclusion,
\begin{equation}
\Omega=\left.\arcsin \left(\frac{v_{\theta}}{v}\right)\right|_{0} \rightarrow \mp \pi / 2  \text { for } \theta \rightarrow 0.
\end{equation}

{\bf B: $\theta\rightarrow\frac{\pi}{2}$}

A similar approach in the  $\theta \rightarrow 0$ limit, after defining the local coordinate as  
\begin{equation}
\rho^2:=g_{\phi_2 \phi_2}.
\end{equation}
yields
\begin{equation}
\sigma_{\pm}\approx \text{constant}\pm \frac{1}{\rho}\times \text{constant}.
\end{equation}
In conclusion
\begin{equation}
\Omega=\left.\arcsin \left(\frac{v_{\theta}}{v}\right)\right|_{\frac{\pi}{2}} \rightarrow \pm \pi / 2  \text { for } \theta \rightarrow \frac{\pi}{2}.
\end{equation}

{\subsubsection {Horizon limit}}

In order to investigate the behavior of the metric components while approaching the horizon, it is useful to introduce the lapse function, defined as
\begin{eqnarray}
&N&:=\left(-g^{tt}\right)^{\left(-\frac{1}{2}\right)}\\
&=&\left[-\frac{g_{t \phi_{2}}^2 g_{\phi_{1} \phi_{1}}-2g_{t \phi_{1}} g_{t \phi_{2}} g_{\phi_{1} \phi_{2}}+g_{t \phi_{1}}^2 g_{\phi_{2} \phi_{2}}}{g_{\phi_{1} \phi_{2}}^2 - g_{\phi_{1} \phi_{1}} g_{\phi_{2} \phi_{2}}}-g_{tt}\right]^{\frac{1}{2}}.\nonumber
\end{eqnarray}
Recalling that 
\begin{eqnarray}
D&=&g_{\phi_1 \phi_1} g_{t \phi_2}^2 + g_{\phi_2 \phi_2} g_{t \phi_1}^2 + g_{tt} g_{\phi_1 \phi_2}^2 \nonumber\\
&&- g_{tt} g_{\phi_1 \phi_1} g_{\phi_2 \phi_2} - 2 g_{t \phi_{1}} g_{t \phi_{2}} g_{\phi_1 \phi_2}
\end{eqnarray}
 one can write the lapse function as 
\begin{eqnarray}
N^2&=&-\frac{D}{g_{\phi_{1} \phi_{2}}^2 - g_{\phi_{1} \phi_{1}} g_{\phi_{2} \phi_{2}}}.
\end{eqnarray}
The Killing vector field (with constant $\Omega_{1,2}$)
\begin{equation}
\chi := \partial_{t} + \Omega_1 \partial_{\phi_{1}} + \Omega_2 \partial_{\phi_{2}}
\end{equation}
is null on the horizon,
\begin{equation}
\left(\chi_{\mu} \chi^{\mu}\right)\arrowvert_{H}=0,
\end{equation}
given that these constants are chosen as
\begin{equation}
\Omega_1:=\left(\frac{g_{t \phi_{2}} g_{\phi_{1} \phi_{2}} - g_{t \phi_{1}} g_{\phi_2 \phi_2}}{g_{\phi_{1} \phi_{1}} g_{\phi_{2} \phi_{2}}-g_{\phi_{1} \phi_{2}}^2}\right)_H
\end{equation}
and
\begin{equation}
\Omega_2:=\left(\frac{g_{t \phi_{1}} g_{\phi_{1} \phi_{2}} - g_{t \phi_{2}} g_{\phi_1 \phi_1}}{g_{\phi_{1} \phi_{1}} g_{\phi_{2} \phi_{2}}-g_{\phi_{1} \phi_{2}}^2}\right)_H.
\end{equation}
By defining two functions that extend $\Omega_{1,2}$ beyond the horizon 
\begin{equation}
\omega_1:=\frac{g_{t \phi_{2}} g_{\phi_{1} \phi_{2}} - g_{t \phi_{1}} g_{\phi_2 \phi_2}}{g_{\phi_{1} \phi_{1}} g_{\phi_{2} \phi_{2}}-g_{\phi_{1} \phi_{2}}^2},
\end{equation}
and
\begin{equation}
\omega_2:=\frac{g_{t \phi_{1}} g_{\phi_{1} \phi_{2}} - g_{t \phi_{2}} g_{\phi_1 \phi_1}}{g_{\phi_{1} \phi_{1}} g_{\phi_{2} \phi_{2}}-g_{\phi_{1} \phi_{2}}^2},
\end{equation}
we can write the effective potential  functions as
\begin{eqnarray}
\sigma_{\pm}
&=&\omega_1+\omega_2\pm N\sqrt{\frac{g_{\phi_1 \phi_1}+g_{\phi_2 \phi_2}-2g_{\phi_1 \phi_2}}{g_{\phi_1 \phi_1} g_{\phi_2 \phi_2}-g_{\phi_1 \phi_2}^2}}.
\end{eqnarray}
Note that this can be expressed as a function of the Killing vectors $\xi, \eta_1,\eta_2$, but the above form is easier to work with. 

Now, we introduce two coordinates, $n, z$, where $n$ represents the normal distance to the horizon which vanishes at the horizon, and the $z$ direction is perpendicular to $n$. One can write the other components as a function of these two coordinates.
Without loss of generality, we can safely assume, outside the horizon, $g_{nn}=1$. By generalizing the approximation ideas introduced in \cite{Medved}, we obtain
\begin{eqnarray}
g_{\phi_1 \phi_1}\left(n,z\right)&=&\left[g_H\right]_{\phi_1 \phi_1}\left(z\right)+O\left(n^2\right), \nonumber \\ 
g_{\phi_2 \phi_2}\left(n,z\right)&=&\left[g_H\right]_{\phi_2 \phi_2}\left(z\right)+O\left(n^2\right),\nonumber \\
g_{\phi_1 \phi_2}\left(n,z\right)&=&\left[g_H\right]_{\phi_1 \phi_2}\left(z\right)+O\left(n^2\right),\nonumber\\
N\left(n,z\right)&=&\kappa_H n +O\left(n^3\right),\nonumber\\
\omega_1\left(n,z\right)&=&\Omega_{1}+O\left(n^3\right),\nonumber\\
\omega_2\left(n,z\right)&=&\Omega_{2}+O\left(n^3\right),
\end{eqnarray}
where $\kappa_H$ is the surface gravity which is a nonzero constant on the horizon. Therefore, near the horizon, the effective potential functions become
\begin{eqnarray}
\sigma_{\pm}&\approx&\Omega_{1}+\Omega_{2}\pm \kappa_H n\nonumber\\
&&\times \sqrt{\frac{\left[g_H\right]_{\phi_1 \phi_1}+\left[g_H\right]_{\phi_2 \phi_2}-2\left[g_H\right]_{\phi_1 \phi_2}}{\left[g_H\right]_{\phi_1 \phi_1}\left[g_H\right]_{\phi_2 \phi_2}-\left[g_H\right]_{\phi_1 \phi_2}^2}}\nonumber\\
&\approx&\text{constant}\pm \kappa_H n \times \text{constant}.
\end{eqnarray}
Since, 
\begin{equation}
\frac{1}{\sqrt{g_{nn}}}\frac{\partial}{\partial_{n}}=\frac{1}{\sqrt{g_{rr}}}\frac{\partial}{\partial_{r}}
\end{equation}
near the horizon, and $g_{nn}=1$, we have 
\begin{eqnarray}
v_{r,\pm}&=&\frac{1}{\sqrt{g_{rr}}}\frac{\partial}{\partial_r}\sigma_{\pm}=\frac{\partial}{\partial_{n}}\sigma_{\pm}\label{horizon_limit}\\
&\approx&\pm\kappa_H \sqrt{\frac{\left[g_H\right]_{\phi_1 \phi_1}+\left[g_H\right]_{\phi_2 \phi_2}-2\left[g_H\right]_{\phi_1 \phi_2}}{\left[g_H\right]_{\phi_1 \phi_1}\left[g_H\right]_{\phi_2 \phi_2}-\left[g_H\right]_{\phi_1 \phi_2}^2}}.\nonumber
\end{eqnarray}
The important observation is that there is no sign change in the horizon limit. Therefore we can say that for positive effective potential, $v_r$ is always positive, and for negative effective potential, $v_r$ is always negative. Nevertheless, the angular component ($v_\theta$) changes sign during the contour, because for a positive rotation sense, it starts in the positive direction and ends in the negative direction. This contributes a negative half winding, as expected.
\\

{\subsubsection {Asymptotic limit}}

In the asymptotic limit, we have a flat spacetime in spherical coordinates
\begin{eqnarray}
&&g_{tt}\approx -1, \hspace{0.7 cm} g_{t\phi_1}=0, \hspace{0.5 cm} g_{t\phi_2}=0,\label{asymptotic} \\
&&g_{rr}\approx 1, \hspace{0.9 cm} g_{\theta \theta}\approx r^2, \hspace{0.5 cm} g_{\phi_{1} \phi_{2}}=0, \nonumber\\
&&g_{\phi_1 \phi_1}\approx r^2 \sin ^2 \theta, \hspace{0.5 cm} g_{\phi_2 \phi_2}\approx r^2 \cos ^2 \theta. \nonumber
\end{eqnarray}
This yields
\begin{eqnarray}
v_{r,\pm}&=&\frac{1}{\sqrt{g_{rr}}}\frac{\partial}{\partial_r}\sigma_{\pm}=\frac{\partial}{\partial_{r}}\sigma_{\pm}
\end{eqnarray}
where
\begin{eqnarray}
\sigma_{\pm}&=&\pm \sqrt{-g_{tt}\left(\frac{1}{g_{\phi_1\phi_1}}+\frac{1}{g_{\phi_2\phi_2}}\right)}\nonumber\\
&=&\pm \frac{1}{r\sin \theta \cos \theta}.
\end{eqnarray}
Therefore,
\begin{eqnarray}
v_{r,\pm}&=&\mp \frac{1}{r^2\sin \theta \cos \theta}.
\end{eqnarray}
Since to $\theta \in \left[0,\frac{\pi}{2}\right]$, $\sin \theta \cos \theta >0$, one has
\begin{equation}
\text{sign}\left(v_{r,\pm}\right)\arrowvert_{\infty}=\mp 1.
\end{equation}
This is sufficient. The angular component changes sign and it contributes a negative half winding. The discussion in the paragraph including (\ref{windingC}) applies here verbatim and one gets a total winding number -1 which refers to a standard light ring.
\\
\\
\subsection{ Case II: Distinct angular momenta}
At this point, we would like to investigate the solutions with distinct angular momenta of light $\Phi_1 \ne \Phi_2$. The effective potential can be written as
\begin{eqnarray}
V&=&-\frac{1}{D}\left(g_{\phi_{1}\phi_{1}}g_{\phi_{2}\phi_{2}}-g_{\phi_{1}\phi_{2}}^2\right)\\
&&\times\left[E^2+E\left(2\Phi_1\left(\frac{g_{t\phi_{1}}g_{\phi_{2}\phi_{2}}-g_{t\phi_{2}}g_{\phi_{1}\phi_{2}}}{g_{\phi_{1}\phi_{1}}g_{\phi_{2}\phi_{2}}-g_{\phi_{1}\phi_{2}}^2}\right)\right.\right.\nonumber\\
&&\left.\left.2\Phi_2\left(\frac{g_{t\phi_{2}}g_{\phi_{1}\phi_{1}}-g_{t\phi_{1}}g_{\phi_{1}\phi_{2}}}{g_{\phi_{1}\phi_{1}}g_{\phi_{2}\phi_{2}}-g_{\phi_{1}\phi_{2}}^2}\right)\right)\right.\nonumber\\
&&\left.+\left(\Phi_1^2\left(\frac{g_{tt}g_{\phi_{2}\phi_{2}}-g_{t\phi_{2}}^2}{g_{\phi_{1}\phi_{1}}g_{\phi_{2}\phi_{2}}-g_{\phi_{1}\phi_{2}}^2}\right)\right.\right.\nonumber\\
&&\left.\left.+\Phi_2^2\left(\frac{g_{tt}g_{\phi_{1}\phi_{1}}-g_{t\phi_{1}}^2}{g_{\phi_{1}\phi_{1}}g_{\phi_{2}\phi_{2}}-g_{\phi_{1}\phi_{2}}^2}\right)\right.\right.\nonumber\\
&&\left.\left.2\Phi_1\Phi_2\left(\frac{g_{t\phi_{1}}g_{t\phi_{2}}-g_{tt}g_{\phi_{1}\phi_2}}{g_{\phi_{1}\phi_{1}}g_{\phi_{2}\phi_{2}}-g_{\phi_{1}\phi_{2}}^2}\right)\right)\right]
.\nonumber
\end{eqnarray}
Here instead of the inverse impact parameters $\sigma_{1,2}$,  it pays to use the impact parameters $b_{1,2}$, hence pulling out the $E^2$, the effective potential factors as
\begin{equation}
V=-\frac{E^2}{D}\left(g_{\phi_{1}\phi_{1}}g_{\phi_{2}\phi_{2}}-g_{\phi_{1}\phi_{2}}^2\right)\left(1-b_+\right)\left(1-b_-\right),
\end{equation}where
\begin{eqnarray}
b_\pm&:=&-b_1\left(\frac{g_{t\phi_{1}}g_{\phi_{2}\phi_{2}}-g_{t\phi_{2}}g_{\phi_{1}\phi_{2}}}{g_{\phi_{1}\phi_{1}}g_{\phi_{2}\phi_{2}}-g_{\phi_{1}\phi_{2}}^2}\right)\label{impact1}\\
&&-b_2\left(\frac{g_{t\phi_{2}}g_{\phi_{1}\phi_{1}}-g_{t\phi_{1}}g_{\phi_{1}\phi_{2}}}{g_{\phi_{1}\phi_{1}}g_{\phi_{2}\phi_{2}}-g_{\phi_{1}\phi_{2}}^2}\right)
\pm\frac{1}{2}\sqrt{\Delta}\,\nonumber 
\end{eqnarray}
and
\begin{eqnarray}
\Delta&=&\frac{4 D}{\left(g_{\phi_{1}\phi_{1}}g_{\phi_{2}\phi_{2}}-g_{\phi_{1}\phi_{2}}^2\right)^2}\label{impact2}\\
&\times&\left(g_{\phi_{2}\phi_{2}}b_1^2+g_{\phi_{1}\phi_{1}} b_2^2-2g_{\phi_{1}\phi_{2}} b_1b_2\right).\nonumber
\end{eqnarray}
From now on, the vector field will be defined as
\begin{equation}
v_r :=\frac{\partial_r b_{\pm}}{\sqrt{g_{rr}}}, \hspace{0.5 cm} v_{\theta}:=\frac{\partial_{\theta} b_{\pm}}{\sqrt{g_{\theta\theta}}}. \label{impact3}
\end{equation}
\subsubsection{Axis limit-I }
One can show that  as $\theta\rightarrow0$, one has
\begin{eqnarray}
D&\approx&\rho^2 \left(\left(g_{t \phi_2}^{\left(0\right)}\right)^2 + g_{\phi_2 \phi_2}^{\left(0\right)} \left(\frac{\rho^{2n}}{\rho^2}\right) + g_{tt}^{\left(0\right)} \left(\frac{\rho^{2m}}{\rho^2}\right)\right. \nonumber\\
&&\left.- g_{tt}^{\left(0\right)} g_{\phi_2 \phi_2}^{\left(0\right)} - 2 g_{t \phi_{2}}^{\left(0\right)} \left(\frac{\rho^{n+m}}{\rho^2} \right)\right)\nonumber\\
&\approx& \rho^2 \left(\left(g_{t \phi_2}^{\left(0\right)}\right)^2- g_{tt}^{\left(0\right)} g_{\phi_2 \phi_2}^{\left(0\right)}\right).
\end{eqnarray}
Therefore,  
\begin{eqnarray}
b_{\pm}&\approx&\frac{1}{\rho^2 g_{\phi_2\phi_2}^{\left(0\right)}-\rho^{2m}}\\
&&\times \left[-b_1\left(\rho^n g_{\phi_2 \phi_2}^{\left(0\right)}-g_{t \phi_2}^{\left(0\right)}\rho^m\right)\right.\nonumber\\
&&\left.-b_2\left(\rho^2 g_{t \phi_2}^{\left(0\right)}-\rho^n\rho^m\right)\right.\nonumber\\
&&\left.\pm\rho\sqrt{\left(g_{t\phi_{2}}^{\left(0\right)}\right)^2-g_{tt}^{\left(0\right)}g_{\phi_{2}\phi_{2}}^{\left(0\right)}}\right.\nonumber\\
&&\left.\sqrt{g_{\phi_{2} \phi_{2}}^{\left(0\right)}b_1^2+\rho^2 b_2^2-2\rho^m b_1 b_2}\right].\nonumber
\end{eqnarray}
Keeping the leading and the next to leading terms, one has 
\begin{eqnarray}
b_{\pm}&\approx&\frac{1}{g_{\phi_{2} \phi_{2}}^{\left(0\right)}}\left[-b_2 g_{t \phi_{2}}^{\left(0\right)}\pm\frac{1}{\rho}\arrowvert b_1\arrowvert\sqrt{g_{\phi_{2} \phi_{2}}^{\left(0\right)}}\right.\nonumber\\
&&\left.\times\sqrt{\left(g_{t\phi_{2}}^{\left(0\right)}\right)^2-g_{tt}^{\left(0\right)}g_{\phi_{2}\phi_{2}}^{\left(0\right)}}\right],\nonumber
\end{eqnarray}
which compactly reads:
\begin{equation}
b_\pm \approx \text{constant}\pm\frac{1}{\rho}\times \text{constant}.
\end{equation}
Therefore, $v_r \propto \frac{1}{\rho}$ and $v_{\theta} \propto \frac{1}{\rho^2}$. Thus, we obtained the same result  as in the equal angular momenta case.
\subsubsection{Axis limit -II}
This time, as $\theta\rightarrow\frac{\pi}{2}$,
\begin{eqnarray}
b_{\pm}&\approx&\frac{1}{\rho^2 g_{\phi_1\phi_1}^{\left(0\right)}-\rho^{2m}}\\
&&\times \left[-b_1\left(\rho^2 g_{t \phi_1}^{\left(0\right)}-\rho^n\rho^m\right)\right.\nonumber\\
&&\left.-b_2\left(\rho^n g_{\phi_1 \phi_1}^{\left(0\right)}-g_{t \phi_1}^{\left(0\right)}\rho^m\right)\right.\nonumber\\
&&\left.\pm\rho\sqrt{\left(g_{t\phi_{1}}^{\left(0\right)}\right)^2-g_{tt}^{\left(0\right)}g_{\phi_{1}\phi_{1}}^{\left(0\right)}}\right.\nonumber\\
&&\left.\sqrt{g_{\phi_{1} \phi_{1}}^{\left(0\right)} b_2^2+\rho^2 b_1^2-2\rho^m b_1 b_2}\right],\nonumber\\
\end{eqnarray}
which again has the dominant terms given as
\begin{eqnarray}
b_{\pm}
&\approx&\frac{1}{g_{\phi_{1} \phi_{1}}^{\left(0\right)}}\left[-b_1g_{t \phi_{1}}^{\left(0\right)}\pm\frac{1}{\rho}\arrowvert b_2\arrowvert\sqrt{g_{\phi_{1} \phi_{1}}^{\left(0\right)}}\right.\nonumber\\
&&\left.\times\sqrt{\left(g_{t\phi_{1}}^{\left(0\right)}\right)^2-g_{tt}^{\left(0\right)}g_{\phi_{1}\phi_{1}}^{\left(0\right)}}\right].\nonumber
\end{eqnarray}
In other words, we obtained a result in the form of 
\begin{equation}
b_\pm \approx \text{constant}\pm\frac{1}{\rho}\times \text{constant}.
\end{equation}
Therefore, $v_r \propto \frac{1}{\rho}$ and $v_{\theta} \propto \frac{1}{\rho^2}$. We obtained the same result  as in the equal angular momenta case.
\subsubsection{Horizon limit}
We can write the effective potential functions as
\begin{eqnarray}
b_\pm&=&\omega_1 b_1 + \omega_2 b_2 \pm \frac{\sqrt{D}}{g_{\phi_1 \phi_1} g_{\phi_2 \phi_2} - g_{\phi_1 \phi_2}^2}\nonumber \\ 
&&\times \sqrt{g_{\phi_2 \phi_2} b_1^2+g_{\phi_1 \phi_1} b_2^2 -2g_{\phi_1 \phi_2} b_1 b_2}. 
\end{eqnarray}
With the help of the lapse function, one has
\begin{eqnarray}
&&b_\pm=\omega_1 b_1 + \omega_2 b_2\\
&&\pm N \sqrt{\frac{g_{\phi_2 \phi_2} b_1^2+g_{\phi_1 \phi_1} b_2^2 -2g_{\phi_1 \phi_2} b_1 b_2}{g_{\phi_1 \phi_1} g_{\phi_2 \phi_2} - g_{\phi_1 \phi_2}^2}}, \nonumber
\end{eqnarray}
which in the near the horizon limit, yields
\begin{eqnarray}
b_\pm&\approx&\Omega_{1} b_1 + \Omega_{2} b_2 \\
&&\pm \kappa_H n \sqrt{\frac{g_{H, \phi_2 \phi_2} b_1^2+g_{H, \phi_1 \phi_1} b_2^2 -2g_{H, \phi_1 \phi_2} b_1 b_2}{g_{H, \phi_1 \phi_1} g_{H, \phi_2 \phi_2} - g_{H, \phi_1 \phi_2}^2}}\nonumber\\
&\approx& \text{constant} \pm \kappa_H n\times \text{constant}.\nonumber
\end{eqnarray}
Since, we have
\begin{equation}
v_{r,\pm}=\partial_{n} b_{\pm}
\end{equation}
the discussion around (\ref{horizon_limit}) for the equal momenta case applies here; and this gives a negative half winding.

\subsubsection{Asymptotic limit}
\begin{figure}
	\centering
	\includegraphics[width=0.8\linewidth]{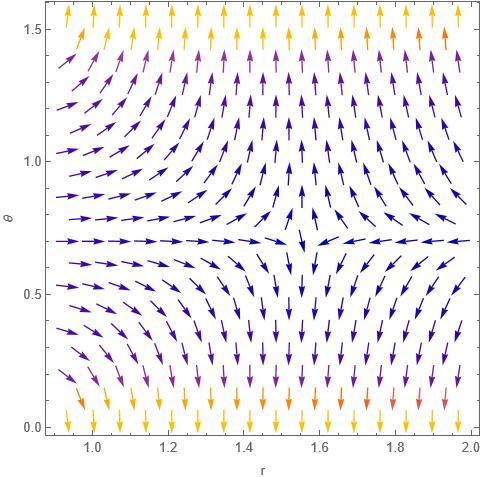}
	\caption{{\bf{Black hole case}} The vector field, $\vec{v}$, obtained by using the impact parameter $b_+$ as given in (\ref{impact1}), (\ref{impact2}), (\ref{impact3}), can be seen in the neighborhood of the standard light ring for the Myers-Perry black hole with two distinct angular momenta. For this plot, we assumed that the mass term is $\mu=1$, the rotation parameters are $a=0.1$ and $b=0.4$ and the angular momenta of the photon are $\Phi_1=1.1$ and $\Phi_2=1.5$. The horizon is located at $r_H=0.83$.}
	\label{fig:spinningdistinctmomenta1}
\end{figure}

\begin{figure}
	\centering
	\includegraphics[width=0.8\linewidth]{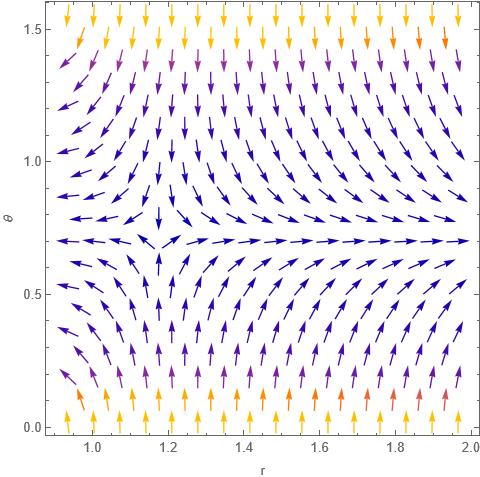}
	\caption{{\bf{Black hole case}} The vector field, $\vec{v}$, obtained by using the impact parameter $b_-$ as given in (\ref{impact1}), (\ref{impact2}), (\ref{impact3}),   can be seen in the neighborhood of the standard light ring for the Myers-Perry black hole with two distinct angular momenta. We took $\mu=1$,  $a=0.1$ and $b=0.4$ and the angular momenta of the photon are $\Phi_1=1.1$ and $\Phi_2=1.5$. The horizon is located at $r_H=0.83$.}
	\label{fig:spinningdistinctmomenta2}
\end{figure}

In the asymptotic limit, one can work in the flat spacetime coordinates, which are shown in (\ref{asymptotic}). This implies that 
\begin{eqnarray}
v_r&=&\frac{1}{g_{rr}}\partial_r b_\pm=\partial_{r} b_\pm. \nonumber
\end{eqnarray}
In this limit,
\begin{eqnarray}
D&\approx&r^4 \sin^2 \theta \cos^2 \theta,
\end{eqnarray}
and therefore
\begin{eqnarray}
\Delta&\approx&\frac{4}{r^2 \sin^2 \theta \cos^2 \theta}\left(\cos^2 \theta b_1^2 + \sin^2 \theta b_2^2\right).
\end{eqnarray}
By using this, we obtain
\begin{eqnarray}
b_\pm&=&\pm \frac{1}{r \sin \theta \cos \theta}\sqrt{\cos^2 \theta b_1^2+ \sin^2 \theta b_2^2}, 
\end{eqnarray} 
and finally calculate
\begin{eqnarray}
v_{r,\pm}&=&\mp\frac{1}{r^2 \sin \theta \cos \theta}
\sqrt{\cos^2 \theta b_1^2+ \sin^2 \theta b_2^2}.
\end{eqnarray}
Since $\theta \in \left[0,\frac{\pi}{2}\right]$, $\sin \theta \cos \theta >0$. In the asymptotic limit, $r\rightarrow\infty$,
\begin{equation}
\text{sign}\left(v_{r,\pm}\right)\arrowvert_{\infty}=\mp 1,
\end{equation}
The angular component changes sign as in the equal momenta case and contributes a negative half winding. 
The general behavior of the vector field around the light ring is plotted in Figs.
\ref{fig:spinningdistinctmomenta1},\ref{fig:spinningdistinctmomenta2}.

\begin{figure}
	\centering
	\includegraphics[width=0.8\linewidth]{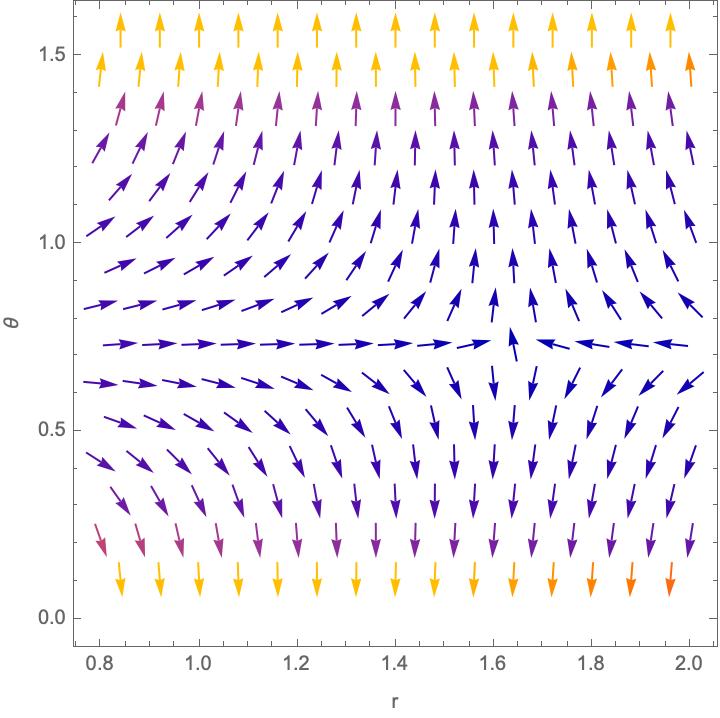}
	\caption{{\bf{No Horizon  case}} The vector field, $\vec{v}$, obtained by using the impact parameter $b_+$ as given in (\ref{impact1}), (\ref{impact2}), (\ref{impact3})  can be seen in the neighborhood of the single light ring for the naked singularity. So this light ring survives. For this plot, we assumed that the mass term is $\mu=1$, the rotation parameters are $a=0.3$ and $b=0.8$ and the angular momenta of the photon are $\Phi_1=1.1$ and $\Phi_2=1.5$. }
	\label{fig:spinningdistinctmomenta3}
\end{figure}
\begin{figure}
	\centering
	\includegraphics[width=0.8\linewidth]{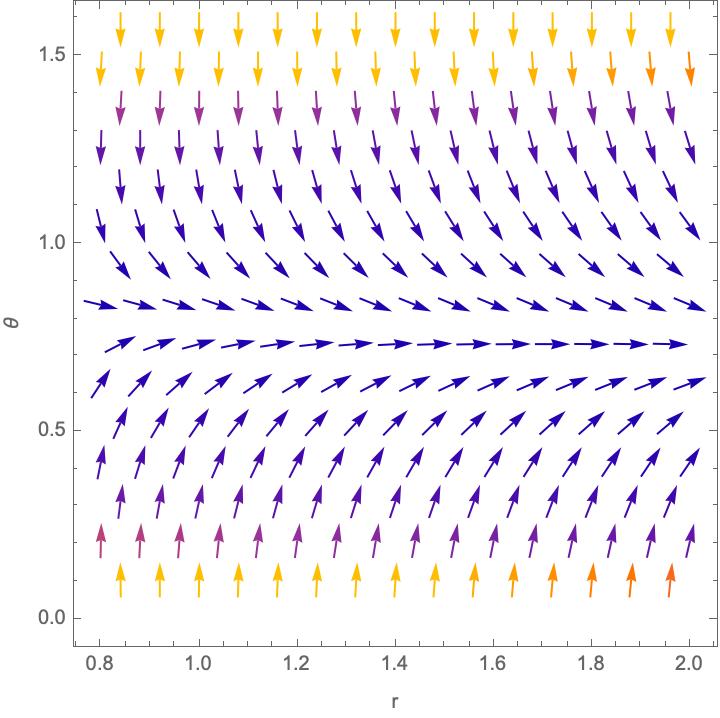}
	\caption{{\bf{No Horizon  case}} The vector field, $\vec{v}$, obtained by using the impact parameter $b_-$ as given in (\ref{impact1}), (\ref{impact2}), (\ref{impact3}) has no winding as  can be seen. The light ring with $b_-$ disappears for the naked singularity case. 
We took $\mu=1$, $a=0.3$ and $b=0.8$ and the angular momenta of the photon are $\Phi_1=1.1$ and $\Phi_2=1.5$.}
	\label{fig:spinningdistinctmomenta4}
\end{figure}

\section{Naked Singularity, Cosmic Censorship and Light Rings}

In the discussions so far, we have assumed the existence of a Killing horizon that hides the singularity in the spacetime as is expected according to the cosmic censorship hypothesis \cite{Penrose}. But, one could ask if this hypothesis can be tested using the light ring structure and the corresponding shadow. This problem will be addressed in full in \cite{Aydin3}. Here we would like to point out that when the horizon disappears, the rotating solution loses one of its light rings as shown in (\ref{fig:spinningdistinctmomenta3},\ref{fig:spinningdistinctmomenta4}): there remains a single light ring around the naked singularity.

 In this work, we have shown this in 4+1 dimensions, but the same phenomenon also appears in the 3+1 dimensional Kerr black hole. Let us show this analytically.
There are two light rings outside the event horizon of a Kerr black (with mass $m$ and the dimensionless rotation parameter $u= a^2/m^2$) which are located in the equatorial plane at the radii ( $r_\pm = m x_\pm$) \cite{Aydin}
\begin{eqnarray}
&& x_\pm = 2 +2\cos \left(\frac{2}{3} \cos ^{-1}(\pm \sqrt{u}  )\right). \label{retro-pro} 
\end{eqnarray}
where $x_+$ is the retrograde orbit satisfying $  3 \le x_+  \le 4$, while $x_{-}$ is the prograde orbit satisfying $  1 \le x_- \le 3$. The event horizon is located at
\begin{equation}
x_{\text{H}}=1+(1-u)^{1/2},
\end{equation}
for $ 0\le u \le 1$ and disappears for larger values of $u$. For any value of $u  >1$, the retrograde orbit $x_+$ remains but the prograde orbit $x_-$ disappears. That means compared to the Kerr black hole with a horizon, the naked singularity has only a single light ring. This could allow us to test the cosmic censorship hypothesis.

\section{Conclusions}

Recent developments \cite{Cunha,Cunha2} suggest that the environment of a black hole (an ultra-compact object with a horizon) has rather unique observable properties which are best studied by light rings. These are null, unstable circular orbits and are closely related to the ringdown and shadow of black holes and the stability of the black hole itself. Ultracompact objects without horizons that could mimic black holes also have light rings at least one of which is stable (proven under certain conditions on the matter forming the ultracompact object \cite{Cunha2}). The stability of the light ring yields a nonlinear instability of the ultracompact object whose fate has been studied for bosonic stars \cite{Cunha3}.

In \cite{Cunha}, topological arguments were established to prove that assuming the existence of a Killing horizon in four-dimensional spacetime, there exists a light ring for each rotation sense for a vacuum axially symmetric, topologically sphere, circular, stationary, asymptotically flat metric. In the current work, we extended these topological arguments to five-dimensional generic stationary black holes without referring to the field equations and showed that there exists a light ring for each rotation sense of the five-dimensional stationary geometry with a Killing horizon. We assumed asymptotic flatness such that far away from the black hole region, the stationary solution reduces to the Myers-Perry spinning solution. We also studied the static case. When applying the topological techniques to five-dimensional black holes, one encounters the difficulty that generically there are two distinct conserved angular momenta for the light which eventually complicates the effective potential: namely, the impact parameters of light enter the effective potential. But these parameters do not spoil the topological arguments in proving the existence of light rings for generic stationary metrics without referring to the underlying field equations. But these parameters do change the location of the light rings in the $r,\theta$ coordinates. Our computations and the examples show that the techniques developed in \cite{Cunha} are robust.

Finally, we have also studied the naked singularity case for which the black hole has no horizon and showed that one of the light rings disappears and one is left with a single light ring. We have also discussed the four-dimensional naked singularity and showed that the prograde light ring of the Kerr black hole disappears while the retrograde orbit is intact. This opens up an interesting discussion regarding a plausible test of the cosmic censorship which will be discussed elsewhere \cite{Aydin3}.

\section{Appendix}
In the first axis limit, $\theta\rightarrow0$ , we assumed that $g_{t\phi_{1}}$ and $g_{\phi_{1} \phi_{2}}$ approach zero faster than or as fast as $g_{\phi_{1} \phi_{1}}$. Here we provide a proof of this assumption. First, we introduce $\rho :=\sqrt{g_{\phi_{1} \phi_1}}$ and $z$ is a coordinate that is orthogonal to $\rho$. The metric can be rewritten as
\begin{eqnarray}
ds^2&=&g_{tt}\left(\rho,z\right)dt^2+2g_{t\phi_1}\left(\rho,z\right)dt d\phi_1\\
&&+2g_{t\phi_2}\left(\rho,z\right)dt d\phi_2+2g_{\phi_1 \phi_2}\left(\rho,z\right)d\phi_1 d\phi_2\nonumber\\
&&+\rho^2 d\phi_1^2+g_{\phi_2 \phi_2}\left(\rho,z\right) d\phi_2^2\nonumber\\
&&+g_{\rho \rho}\left(\rho,z\right)d\rho^2+g_{zz}\left(\rho,z\right)dz^2\nonumber.
\end{eqnarray}
Around the axis, $\rho\rightarrow0$, the metric components can be expanded as
\begin{eqnarray}
g_{tt}&\approx&-1+O\left(\rho\right),\,\,\,
g_{t\phi_2}\approx1+O\left(\rho\right),\nonumber\\
g_{\rho \rho}&\approx&1+O\left(\rho\right),\,\,\,
g_{zz}\approx1+O\left(\rho\right),\nonumber\\
g_{\phi_2 \phi_2}&\approx&1+O\left(\rho\right),\nonumber\\
g_{t\phi_1}&\approx&g_{t\phi_{1}}^{\left(1\right)}\left(z\right)\rho+g_{t\phi_{1}}^{\left(2\right)}\left(z\right)\rho^2+O\left(\rho^2\right),\nonumber\\
g_{t\phi_2}&\approx&g_{t\phi_{2}}^{\left(1\right)}\left(z\right)\rho+g_{t\phi_{2}}^{\left(2\right)}\left(z\right)\rho^2+O\left(\rho^2\right).
\end{eqnarray}
The scalar curvature in the first order of expansion can be computed to be
\begin{equation}
R\approx\frac{g_{t\phi_{1}}^{\left(1\right)}\left(z\right)^2-2 g_{t\phi_{1}}^{\left(1\right)}\left(z\right) g_{t\phi_{2}}^{\left(1\right)}\left(z\right)-g_{t\phi_{2}}^{\left(1\right)}\left(z\right)^2}{2 \rho^2 \left(g_{t\phi_{1}}^{\left(1\right)}\left(z\right)^2-2 g_{t\phi_{1}}^{\left(1\right)}\left(z\right) g_{t\phi_{2}}^{\left(1\right)}\left(z\right)-g_{t\phi_{2}}^{\left(1\right)}\left(z\right)^2+2\right)}.\nonumber 
\end{equation}
For $R$ to be finite as $\rho\rightarrow0$, one must have $g_{t\phi_{1}}^{\left(1\right)}\left(z\right)=0$ and $g_{t\phi_{2}}^{\left(1\right)}\left(z\right)=0$. This completes our proof. The same procedure can be followed for the second axis limit, $\theta\rightarrow\frac{\pi}{2}$, and we reach the conclusion that $g_{t\phi_{2}}$ and $g_{\phi_2 \phi_2}$ approach to zero faster than or as fast as $g_{\phi_2 \phi_2}$.

\end{document}